\documentclass[aps,preprint,showpacs]{revtex4}
\usepackage{graphicx,epsfig}
\usepackage{amssymb}
\begin{document}

\title{Scattering sections from regular black holes immersed in perfect fluid dark matter}

\author{Omar Pedraza}
\email{omarp@uaeh.edu.mx}
\author{L. A. L\'opez}
\email{lalopez@uaeh.edu.mx (corresponding-author)}
\author{Isaac Fern\'andez}
\email{fe445676@uaeh.edu.mx}

\affiliation{ \'Area Acad\'emica de Matem\'aticas y F\'isica, UAEH, 
Carretera Pachuca-Tulancingo Km. 4.5, C P. 42184, Mineral de la Reforma, Hidalgo, M\'exico.}

\begin{abstract}
In this contribution, we investigate the scattering cross sections of black holes immersed in perfect fluid dark matter (PFDM). We present both the classical and semi-classical scattering cross sections for different values of the parameter that characterizes the PFDM contribution. Our results show that the presence of dark matter increases the classical scattering cross section and modifies the width of the interference fringes in the semi-classical regime. In addition, the scattering cross section is also computed using the partial wave method for the black holes considered, exhibiting similar qualitative behavior. These findings suggest that the effects of dark matter surrounding black holes may play an important role in black holes phenomenology, particularly in certain regions near the black hole.
\end{abstract}

\maketitle

\section{Introduction}

In the study of the various properties of black holes (BHs), the information obtained is usually expressed in terms of the different parameters that characterize these objects, such as mass, charge, angular momentum, and more recently, parameters associated with black holes immersed in dark matter or dark energy.
%and angular momentum, as well as, more recently, parameters associated with black holes immersed in dark matter or dark energy.  
For example, several studies have been developed in: geodesic trajectories \cite{Lungu:2025iri} \cite{Heydari-Fard:2022jdu}, quasi-normal modes \cite{Lopez:2023ixb} \cite{Hamil:2024neq}, and the red shift of light emitted by particles orbiting a black hole \cite{Trad:2025xyi}, among others.

To determine how these parameters affect the physical properties of black holes, it is common to analyze test fields propagating in their vicinity. One of the standard approaches is the study of scattering processes.

As a first approximation, scattering phenomena can be investigated through the analysis of geodesic trajectories propagating from infinity toward the black hole and being deflected by an effective potential. Within this framework, one can obtain the classical scattering cross section, as discussed, for example, in \cite{Ramirez:2021ibk},\cite{,Macedo:2015qma}. Moreover, the impact parameter obtained in the classical analysis plays a crucial role in determining the black hole shadow in this approximation, as is shown in \cite{Pedraza:2020uuy}.

However, wave scattering by black holes gives rise to diffraction effects that cannot be captured within a purely geodesic analysis. In this context, the semi-classical or Glory approximation accounts for wave interference effects at large scattering angles, assuming high-frequency planar scalar waves, this approach, for example, is addressed in \cite{Magalhaes:2022pgl}. It has been shown that the widths of the interference fringes are modified by variations in the different parameters characterizing black holes.

Now if we consider the massless planar scalar waves, the analysis starts from the Klein-Gordon equation. After taking into account the background space-time geometry, the problem can be reduced to a Schrödinger-like equation, which must be solved numerically to obtain the scattering cross section.

For the above mentioned, and with the aim of extending previous studies \cite{Tovar:2025apz} on black holes immersed in PFDM, in this work we propose to study the scattering cross sections of different black hole models in order to analyze how the parameter associated with dark matter modifies the behavior of the corresponding cross sections.

The paper is organized as follows. In Sec. II, we present a brief overview of regular black holes immersed in PDFM. In Sec. III, we derive the expressions for the classical and semi-classical scattering cross sections. In Sec. IV, we provide a brief explanation of the determination of the scattering cross section for massless planar scalar waves and analyse the corresponding scattering cross sections. Finally, the conclusions are presented in Sec. V.

\section{Regular black holes immersed in perfect fluid dark matter}

The perfect fluid dark matter  model was originally proposed by Kiselev \cite{Kiselev:2003ah}, and it has motivated extensive research aimed at deriving black hole solutions immersed in PFDM (see, for example \cite{Das:2020yxw} \cite{Abbas:2023pug}). These solutions are characterized by the presence of an additional term in the metric function, which grows logarithmically with the radial coordinate.

Now, then let us consider the line element of spherically symmetric space-time 

\begin{equation}\label{mfa}
	ds^2=-f(r)dt^2+\frac{dr^2}{f(r)}+r^2d\theta^2+r^2\sin^2\theta d\phi^2\,
\end{equation}
where the metric function can be given by 
\begin{equation}\label{ec.rfc}
	f(r)=1-\frac{2m(r)}{r}\,.
\end{equation}

Each specific form of $m(r)$ represents a different black hole solution. In this work, we consider the black hole solutions presented in \cite{Tovar:2025apz}.

For the Hayward BH \cite{Hayward:2005gi} immersed in PFDM, we have; 
\begin{equation}\label{key}
m(r)=\frac{Mr^3}{r^3+2M_{1}\epsilon^2} + \frac{\alpha}{2}\ln\left(\frac{r}{\left|\alpha\right|}\right) \,,
\end{equation} 
similarly, for the Bardeen solution \cite{Bardeen1} immersed in PFDM, the mass function can be express as;
\begin{equation}\label{key}
	m(r)=\frac{M r^3}{\left(r^2+g^2\right)^{3/2}}  + \frac{\alpha}{2}\ln\left(\frac{r}{\left|\alpha\right|}\right)  \,,
\end{equation} 
and for the
%when refering to
Ayon-Beato-Garc\'ia (ABG) \cite{Ayon-Beato:1998hmi} immersed in PFDM, the function (\ref{ec.rfc}) can be written as
\begin{equation}\label{key}
	m(r)=\frac{M r^3}{\left(r^2+q^2\right)^{3/2}}
	-\frac{q^2r^3}{2\left(r^2+q^2\right)^2} + \frac{\alpha}{2}\ln\left(\frac{r}{\left|\alpha\right|}\right) 	\,.
\end{equation} 

Here, $M$ denote the mass parameters of each black hole, $\epsilon$ is a parameter related to the cosmological constant, $g$ represents the magnetic charge, $q$ denotes the electric charge and  the parameter $\alpha$ is associated with the intensity of the copuling with perfect fluid dark matter. 
The energy--momentum tensor of the PFDM is given by $T^{\nu}_{\mu}=\mathrm{Diag}(-\rho, p_{r}, p_{\theta}, p_{\phi})$, where $\rho$ denotes the energy density and $(p_{r}, p_{\theta}, p_{\phi})$ represent the radial and angular pressures. 
The dark matter energy density and pressures depend on the parameter $\alpha$ \cite{Zhang:2020mxi} as follows:

\begin{equation}
-\rho=p_{r}=\alpha / 8 \pi r^{3}  ;\;\;\;\;\;\ p_{\theta}=p_{\phi}=-\alpha/ 8 \pi r^{3}
\end{equation}

It is important to note that, depending on how the energy--momentum tensor is incorporated into the Einstein equations \cite{Das:2021otl}, the sign of the parameter $\alpha$ can change in such a way that the weak energy condition is satisfied and this would then result in a completely different black hole solution.

\begin{figure}[ht]
\begin{center}
\includegraphics [width =0.45 \textwidth ]{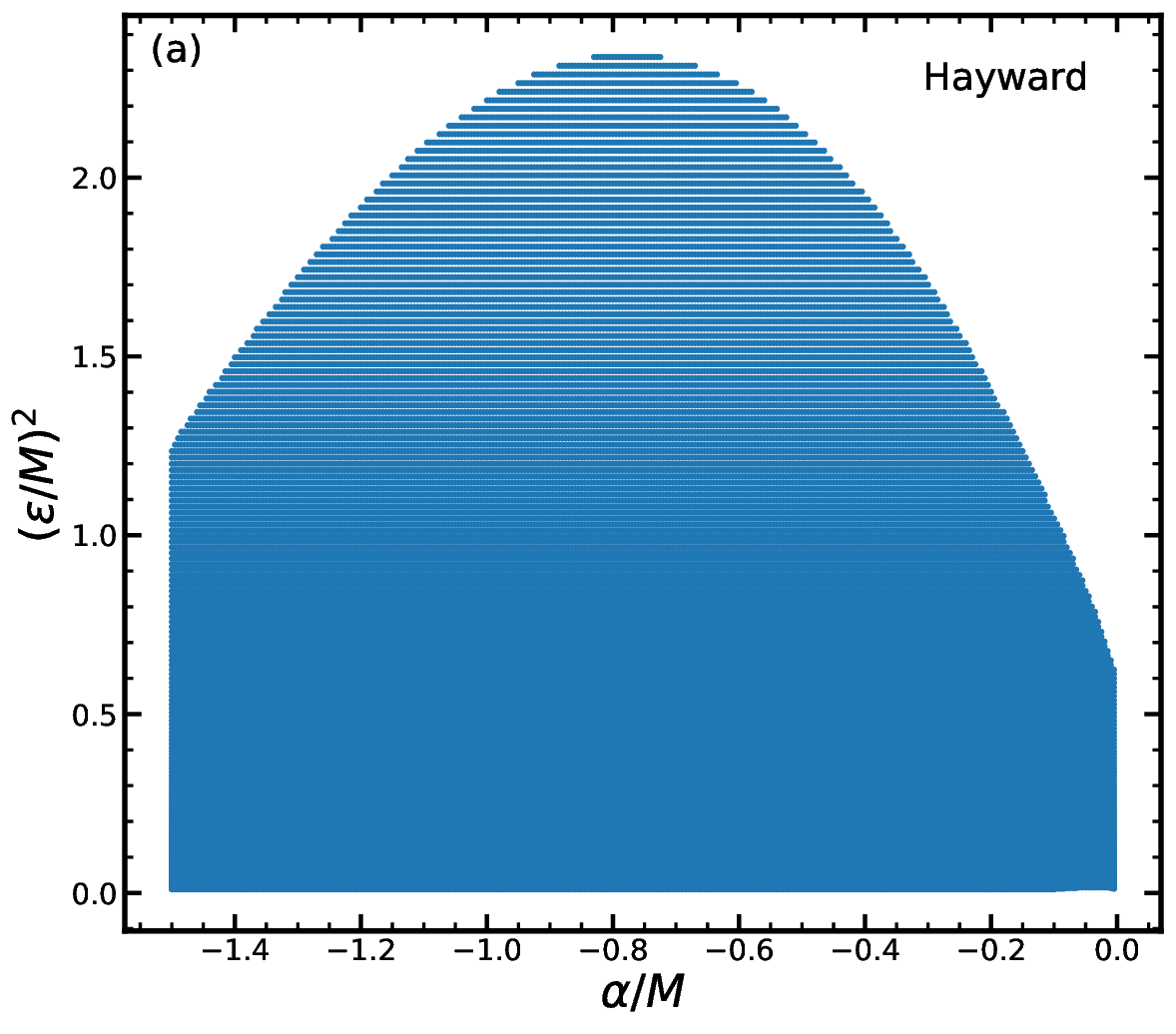}
\includegraphics [width =0.45 \textwidth ]{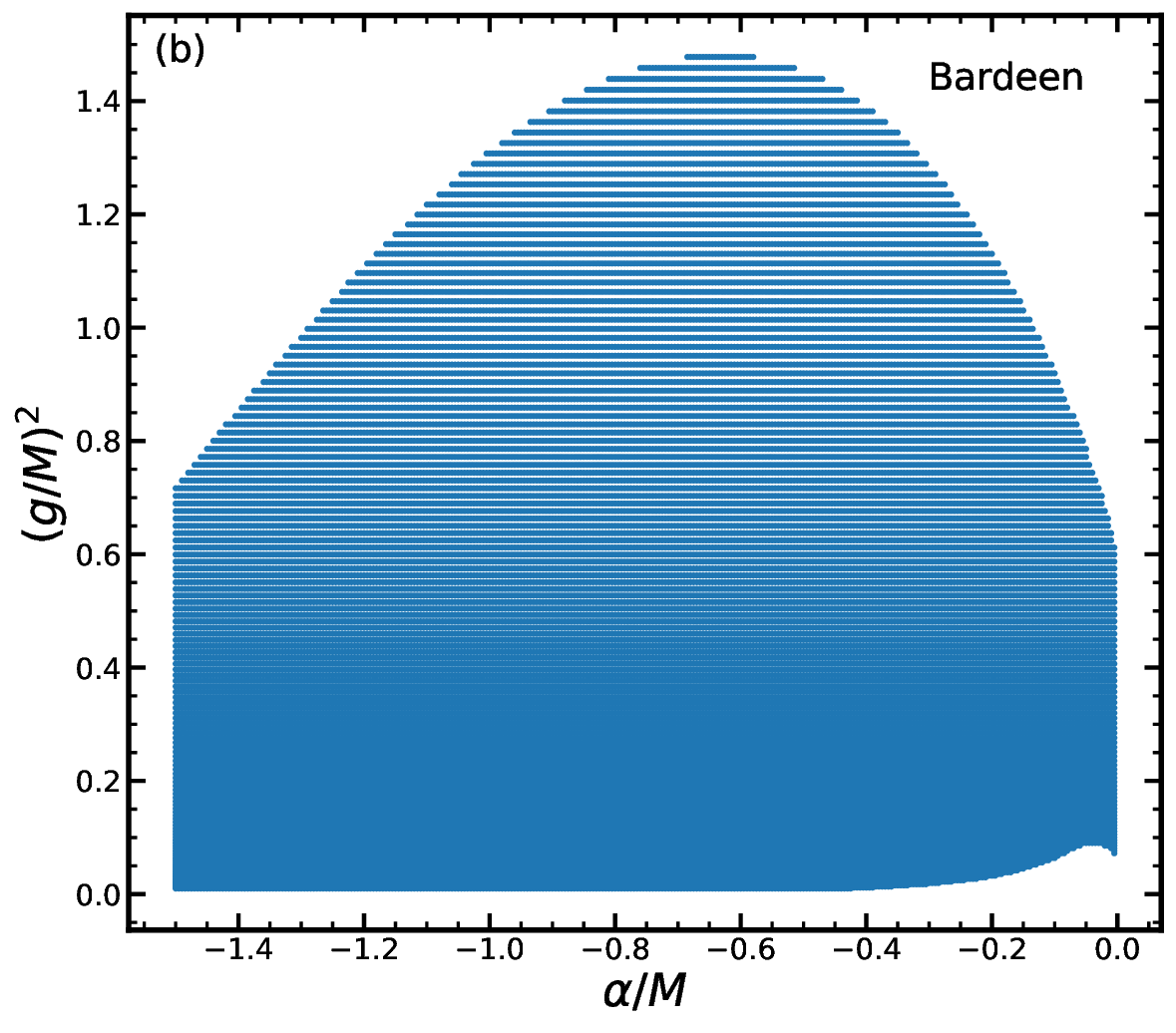}
\includegraphics [width =0.45 \textwidth ]{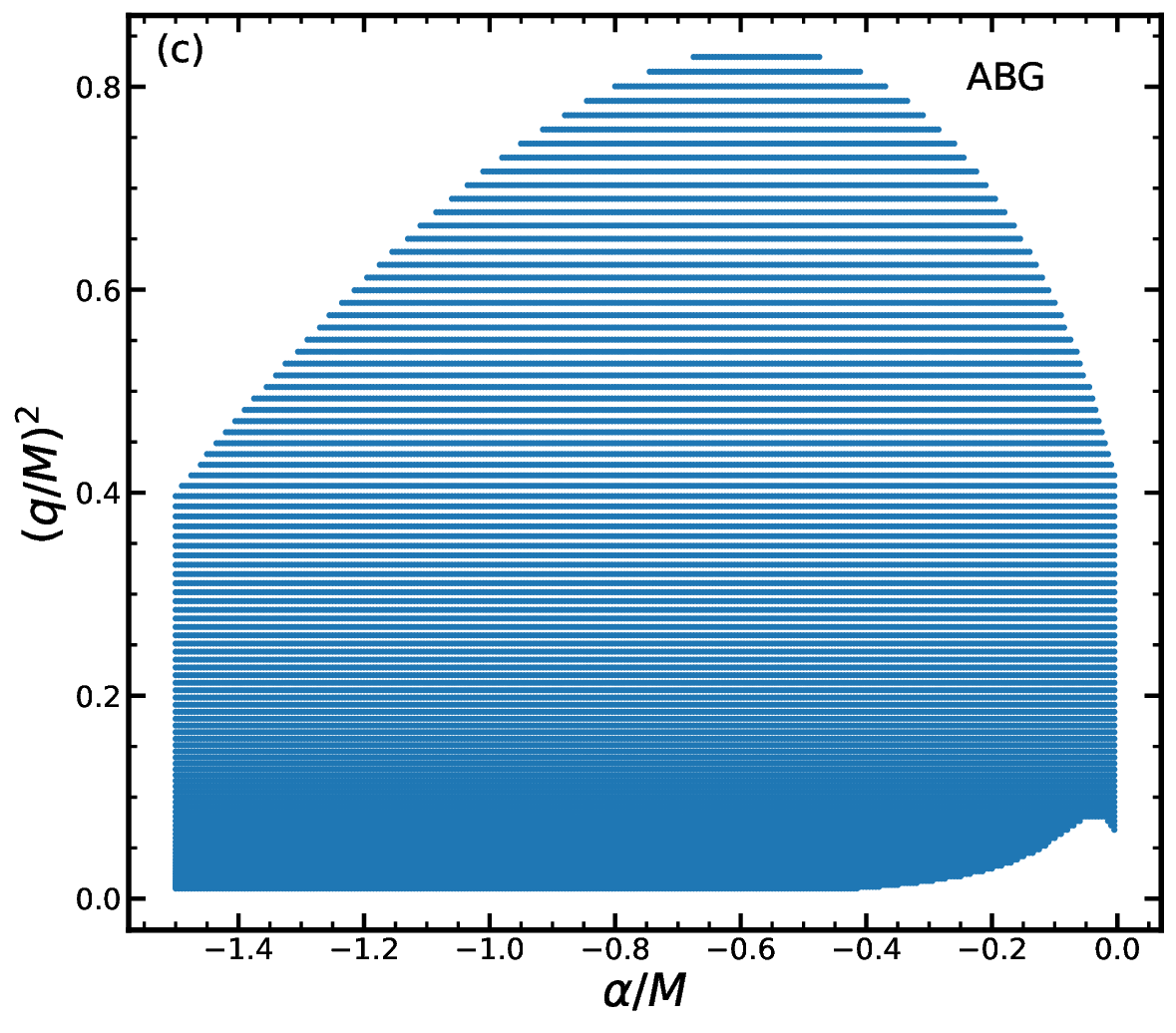}
\includegraphics [width =0.45 \textwidth ]{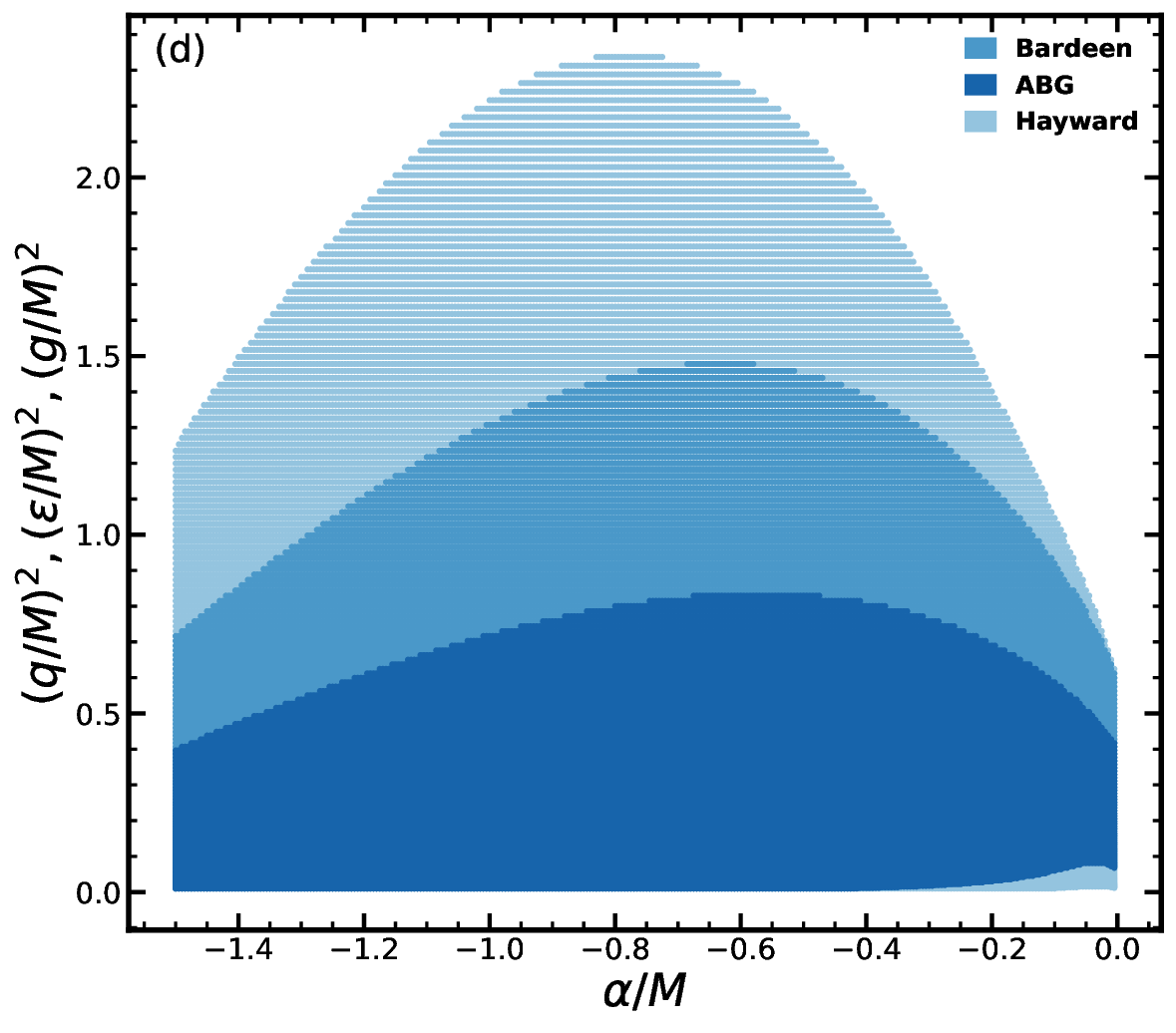}
\end{center}
\caption{Density plots of the parameter ranges for which the Hayward black hole (a), the Bardeen black hole (b), and the Ayón–Beato–García black hole (c) immersed in PFDM exhibit two event horizons. The darker regions indicate the combinations of parameters for which the BHs admits two horizons. Panel (d) shows a comparison of the corresponding regions for the three BH models.
}\label{Fig1}
\end{figure}

The event horizons are defined by the positive roots of the metric function $f(r)$, given in (\ref{ec.rfc}).  We observe that these roots depend strongly on the black hole parameters.  For this analysis we express radial distance and the remaining parameters in units of mass: $r \rightarrow r/M$, $\alpha\rightarrow\alpha/M, \text{ }g\rightarrow g/M,\text{ }q\rightarrow q/M\text{ and } \epsilon\rightarrow \epsilon/M$. The results of the numerical analysis are shown in Fig. (\ref{Fig1} (a), (b), (c)).  The density plots illustrate the combinations of these parameters that lead to the existence of two horizons, which correspond to the darker regions in the different diagrams.
In Fig. (\ref{Fig1} (d)), a comparison is performed, showing that the Hayward black hole exhibits a larger region of allowed parameter values compared to the other BHs.

\section{Classical and Semi-clasical Scattering cross sections}

To obtain the scattering cross section, one commonly employs the study of geodesics within the classical approximation. Geodesic analysis is particularly relevant because, in the high-frequency limit, wave propagation effectively follows the null geodesics behavior \cite{Collins_1973}.

Test particles propagating along null geodesics are described by  $\dot{x}^{\mu}\dot{x}_{\mu} = 0$, where $x^{\mu}=(t,r,\theta,\phi)$ and the overdot indicates the derivative relative to the affine parameter $\tau$.
%differentiation along the affine parameter $\tau$ {\color{blue}Se puede cambiar por }.

For the line element of a spherically symmetric space-time (\ref{mfa}), the conserved quantities associated with the motion of null geodesics are the energy $E$ and the angular momentum $L$, which are given by;

\begin{equation}
E = f(r)\dot{t}, \qquad  L = r^{2}\sin \theta \dot{\phi}
\end{equation}

Now if we restricting the motion to the equatorial plane $\theta = \pi/2$ and solve for $\dot{r}^{2}$, we obtain

\begin{equation} \label{Vef}
\dot{r}^2=E^2-V_{eff},  
\end{equation} where $ V_{eff}$ is the effective potential denoted as $V_{eff} = \frac{L^{2}}{r^{2}}f(r)$.

Since the impact parameter can be defined as $b = L/E$, and we can determine the critical impact parameter $b_{c}$ associated with the unstable circular null orbits $r_{c}$. The critical impact parameter is obtained from (\ref{Vef}) by considering;
\begin{equation}
    \left(\frac{1}{r^2}\frac{\partial r}{\partial \phi}\right)^2 |_{r_{c}}= \frac{1}{b^2}-\frac{1}{r_{c}^2} f(r_{c})=0,
    \label{P}
\end{equation}
so that for any $b > b_{c}$, scattering does not occur.

When we consider geodesics coming from the infinity to a turning point, the deflection angle is given by;

\begin{equation}\label{Theta}
\theta\left(b\right)=2\phi \left( b\right)-\pi\,.
\end{equation}

Following the considerations above, it is then possible to obtain the quantity known as the differential scattering cross section (\ref{Csec}).

\begin{equation}\label{Csec}
   \frac{d\sigma}{d \Omega} = \frac{1}{sin\theta} \sum b(\theta) \left|\frac{db}{d\theta}\right|.
    \label{seccion}
\end{equation}

The summation reflects the possibility that the geodesic becomes temporarily trapped, orbiting around the scattering center before ultimately being deflected.

To obtain the scattering cross section, one must solve (\ref{P}) to establish the relationship between the deflection angle $\theta$ and the impact parameter $b$. A convenient change of variable can be introduced to simplify the numerical evaluation of this equation, leading to
\begin{equation}
    \phi = \int_{0}^{1} 
    \frac{dv}{
    \sqrt{\frac{r_{0}^{2}}{4b^{2}v^{2}}
    - \frac{(1 - v^{2})^{2}}{4v^{2}}\, f(v)}},
    \qquad
    v = \sqrt{1 - \frac{r_{0}}{r}},
    \label{INT2.e}
\end{equation}
where $r_{0}$ denotes the turning points of the effective potential (\ref{Vef}). These turning points lie within the interval $r_{0} \in [r_{c}, \infty)$, where $r_{c}$ is the radius of the closest approach associated with the unstable circular null orbits.

Following the methodology, the value of $r_{c}$ is obtained by considering $V_{\text{eff}}(r_{c})=E$ and $ V'_{\text{eff}}(r_{c}) = 0$, which determine the maximum approach distance prior to the scattering process. Each turning point $r_{0}$ uniquely determines a corresponding value of the impact parameter $b$, establishing a one-to-one relationship between these quantities.

The integral (\ref{INT2.e}) can be evaluated using the Gauss quadrature method. Since this method is iterative, a numerical loop was implemented to run over each pair of values $(r_{0}, b)$, allowing us to construct the relation $\phi(b)$. Substituting this result into in (\ref{Theta}), one then obtains the deflection angle $\theta(b)$, which is required to compute the scattering cross section.

\begin{figure}[ht]
\begin{center}
\includegraphics [width =0.45 \textwidth ]{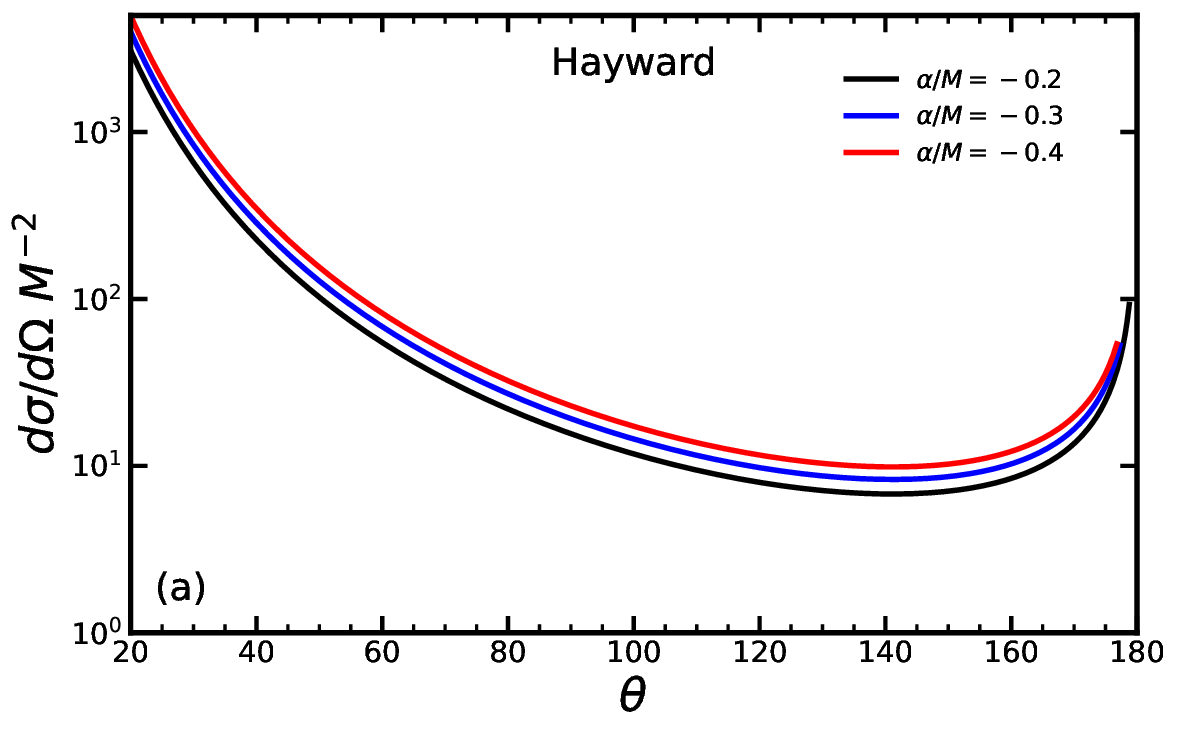}
\includegraphics [width =0.45 \textwidth ]{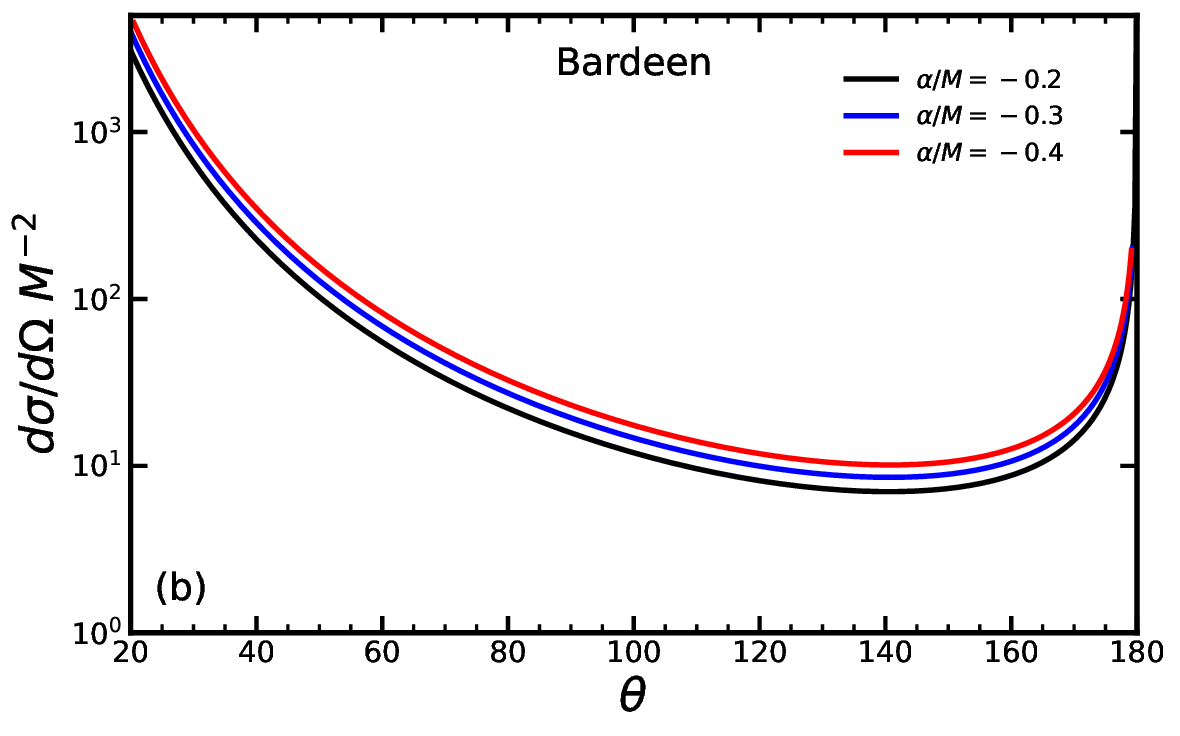}
\includegraphics [width =0.45 \textwidth ]{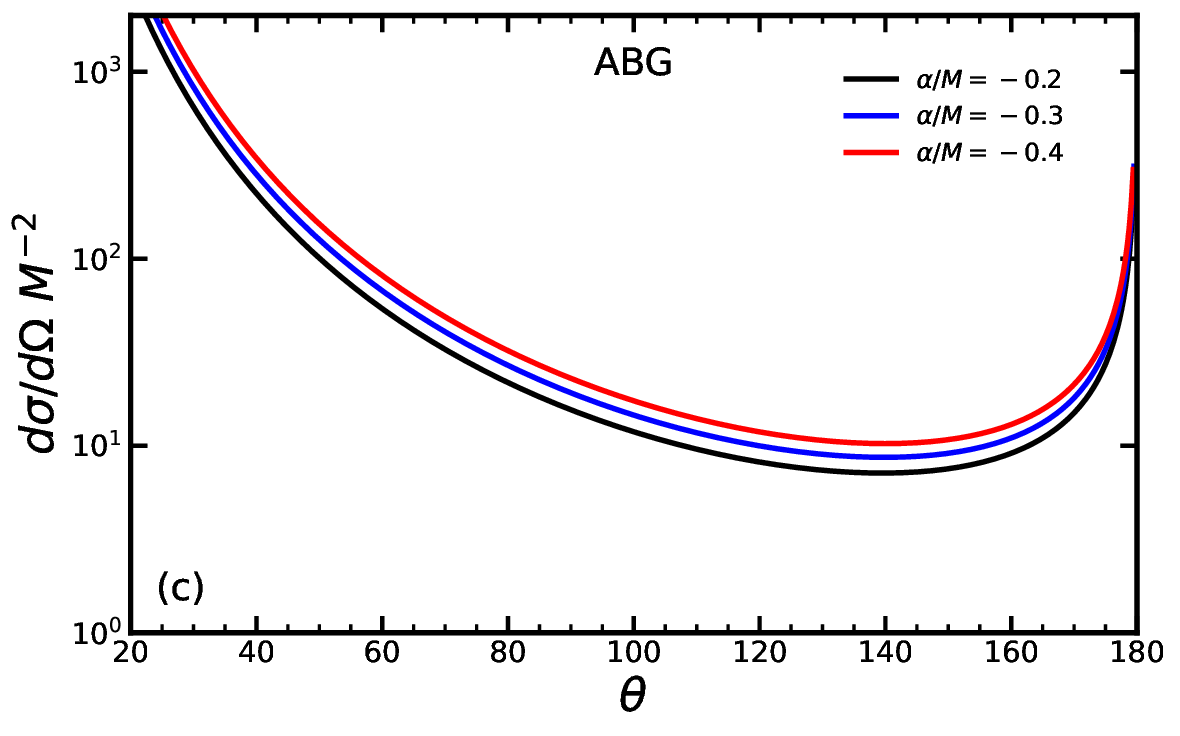}
\includegraphics [width =0.45 \textwidth ]{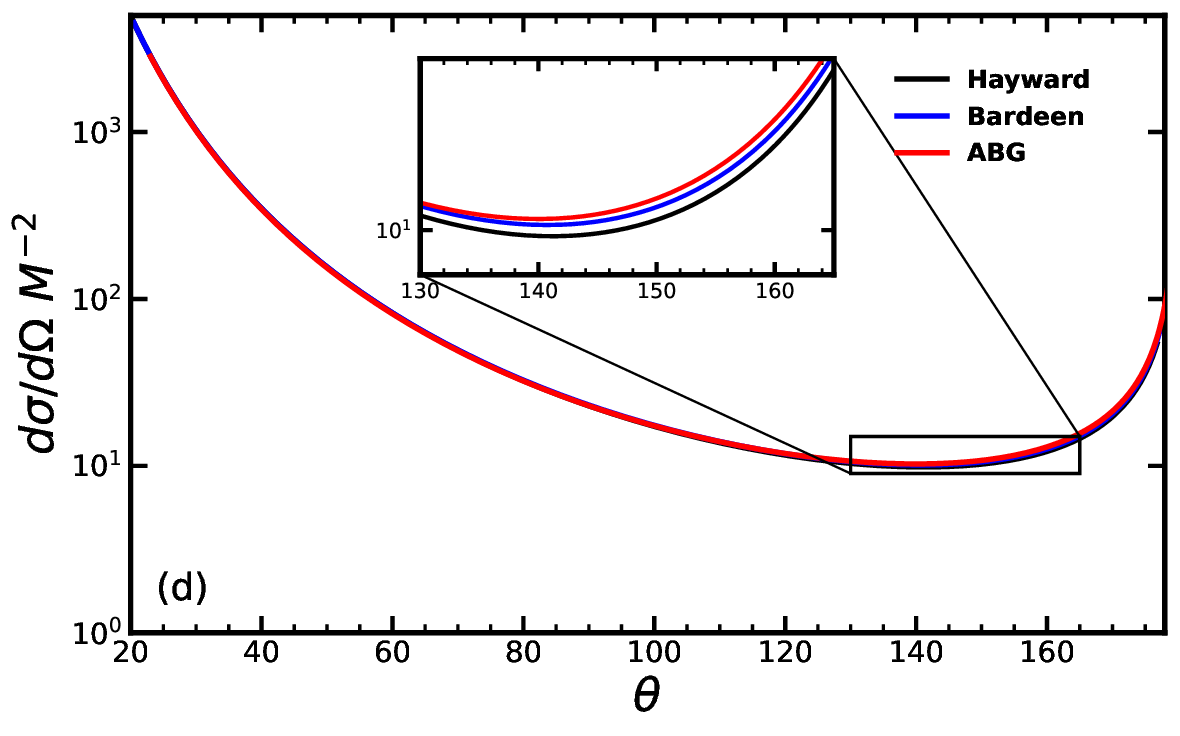}
\end{center}
\caption{Classical scattering cross sections for black holes immersed in PFDM: (a) Hayward black hole, (b) Bardeen black hole and (c) Ay\'on–Beato–Garc\'ia black hole are shown for different values of the parameter $\alpha/M$. Panel (d) presents a comparative analysis of the scattering cross sections for the BHs with $\alpha/ M = -0.4$. For the different panels, we consider $(\epsilon/M)^{2}=(g/M)^{2}=(q/M)^{2}=0.6$}\label{Fig2}
\end{figure}

In Fig. \ref{Fig2}, the classical scattering  cross sections of the different black hole  immersed in PFDM are compared. It can be observed that increasing the dark matter contribution leads to an increase in the scattering cross section in all cases (see Fig.  \ref{Fig2} (a), (b), and (c)). Moreover, no significant differences are found for small and large scattering angles.
By comparing the classical scattering cross sections, we find that the ABG black hole exhibits a larger scattering cross section than the Hayward and Bardeen cases (see Fig.  \ref{Fig2} (d))

 \subsection{ Semi-clasical scattering cross sections}
 
Now when we consider partial waves in the scattering phenomenon, it is necessary to consider the interference that occurs between partial waves with different angular momenta. This situation is not considered by the classical scattering cross section (\ref{seccion}). The approximate method that considers the interference of the waves for low angles and high--frequency scalar plane waves ($\omega \gg 1$) is the semi--classical approach (Glory scattering) \cite{PhysRevD.31.1869}.  The glory approximation of the scalar scattering cross section by spherically symmetric Black Holes is given by;

\begin{equation}\label{Sglory}
\frac{d\sigma_g}{d\Omega}=2\pi w \hat b_g^2 \left| \frac{d\hat b}{d\theta}\right|_{\theta=\pi}J_{2s}^2(\omega \hat b_g \sin\theta)\,,
\end{equation}

with $w$ as the wave frequency, $J_{2s}^2$ as the first kind Bessel function of order $2s$, where $s$ represents the spin ($s=0$ for scalar waves) and the impact parameter of the reflected waves ($\theta \sim\pi$) is denoted by $\hat b_g$. As a semi--classical approximation, it is valid for $M\omega \gg 1$ ($M$ the mass of the BH).

Knowing the glory scattering formula (\ref{Sglory}), it is now only necessary to determine the parameters $b_g$ and the derivative evaluated at $\theta = \pi$. To this end, some of the functions used in the calculation of the classical scattering cross section can be reused. In particular, the value of the impact parameter $b$ corresponding to $\theta = \pi$ can be obtained, and the derivative can be evaluated from the interpolation at $\theta = \pi$. By fixing $M \omega = 2$ and considering a scalar field with particle spin $s = 0$, the differential glory scattering cross section can be obtained by sweeping over the scattering angle $\theta$. 

\begin{figure}[ht]
\begin{center}
\includegraphics [width =0.45 \textwidth ]{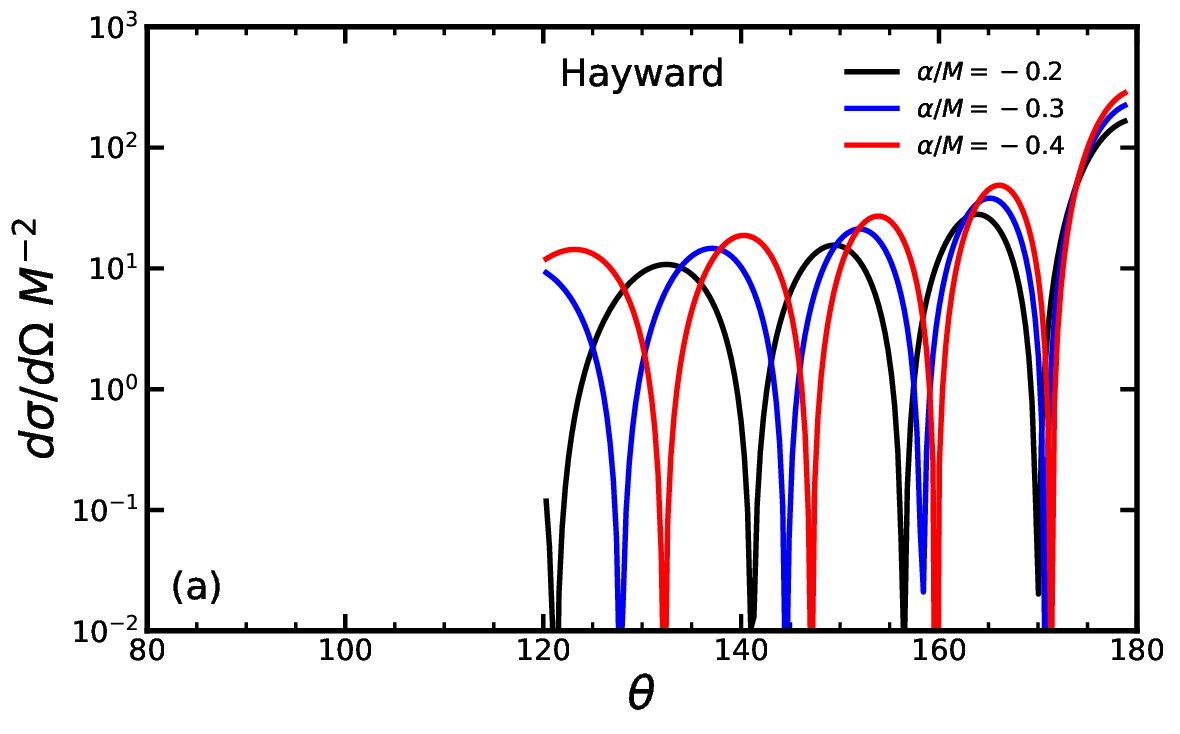}
\includegraphics [width =0.45 \textwidth ]{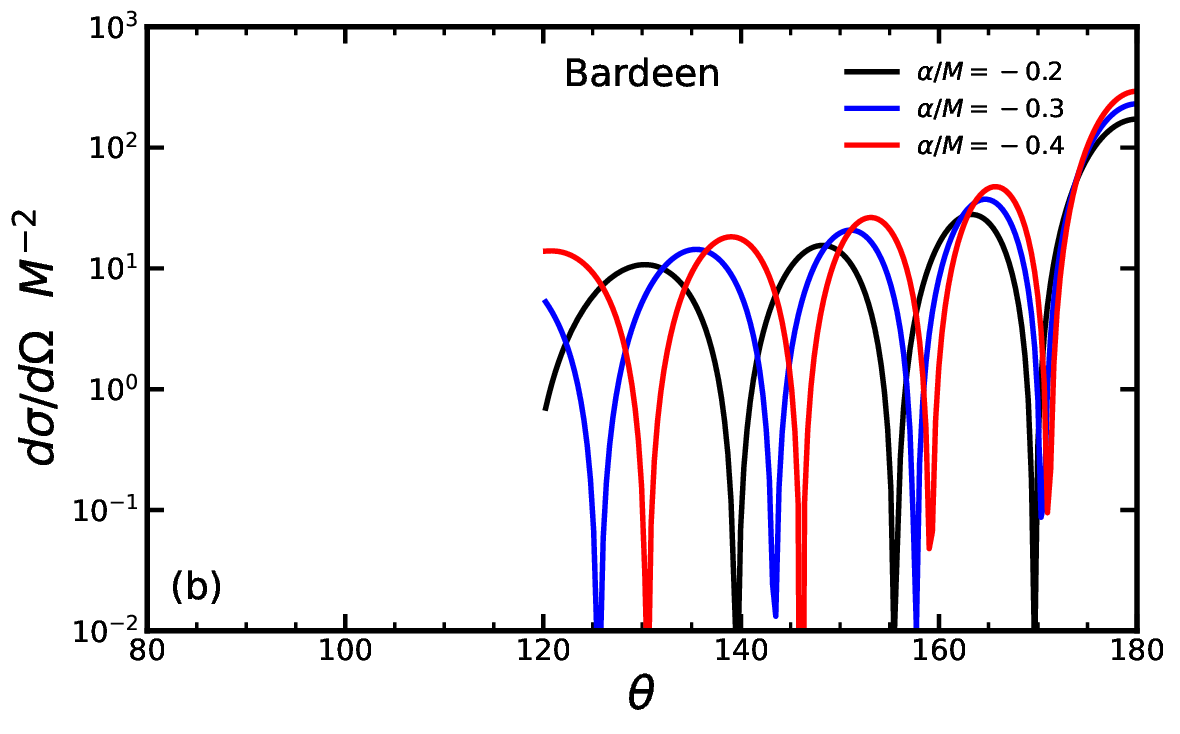}
\includegraphics [width =0.45 \textwidth ]{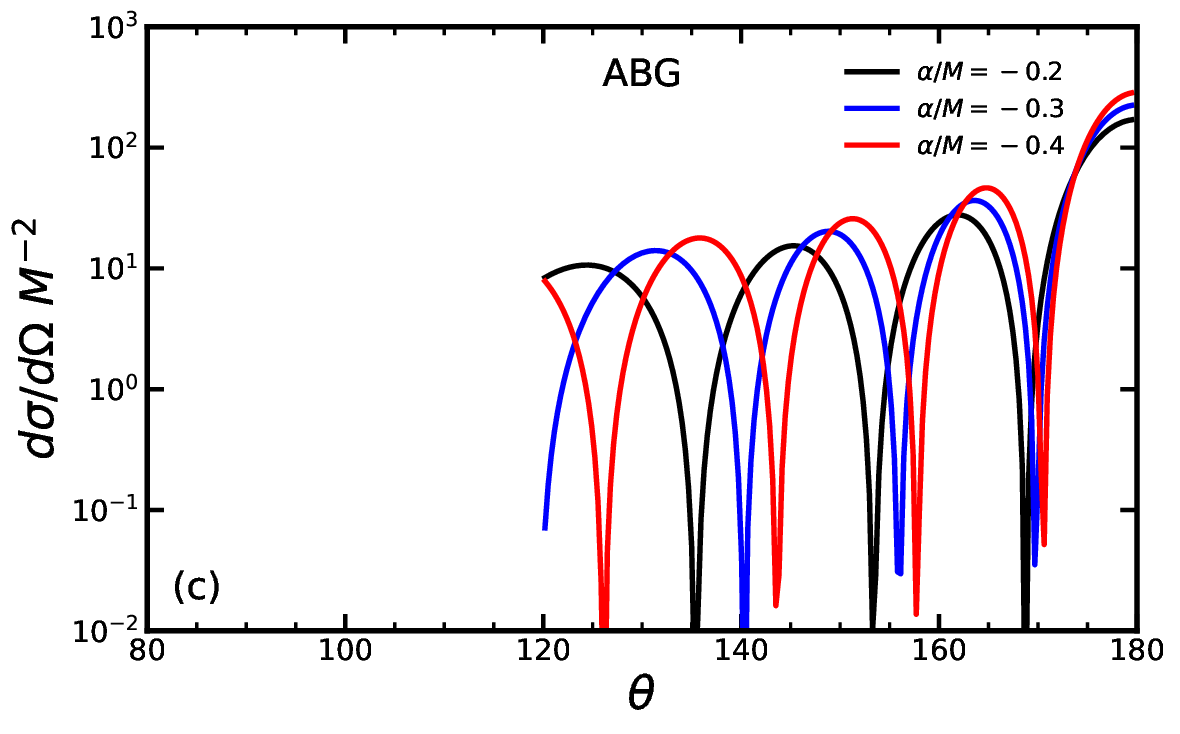}
\includegraphics [width =0.45 \textwidth ]{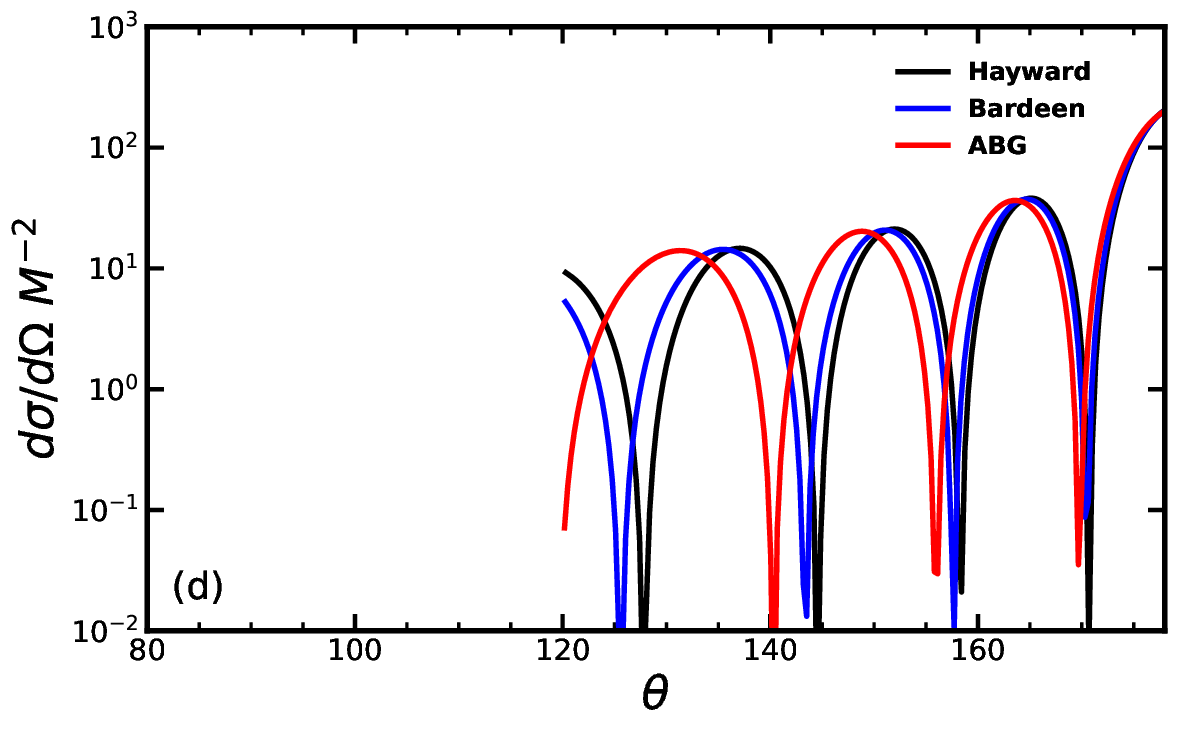}
\end{center}
\caption{Semi-classical scattering cross sections for black holes immersed in PFDM: (a) Hayward black hole, (b) Bardeen black hole, and (c) Ay\'on–Beato–Garc\'ia black hole are shown for different values of $\alpha / M$. Panel (d) presents a comparative analysis of the  semi-classical scattering cross sections for the BHs with $\alpha / M = -0.4$. For the different panels, we consider $(\epsilon/M)^{2}=(g/M)^{2}=(q/M)^{2}=0.6$ and $M\omega=2$}\label{Fig3}
\end{figure}

The semi-classical scattering cross sections for the different BHs are shown in Fig. \ref{Fig3} ((a), (b), and (c)). We observe that the scattering cross section exhibits a qualitatively similar behaviour for different values of the parameter $\alpha$. Nevertheless, it can be inferred that the interference produced by partial scattering waves with different angular momenta within the glory approximation is modified by the presence of this parameter. In particular, the widths of the interference fringes decrease as the contribution associated with the parameter $\alpha$ increases.
It is also worth mentioning that, in general, both in the classical and semi-classical scattering cross sections, the presence of PFDM leads to an increase in the magnitude of the scattering cross sections. Although the scattering cross sections are similar throughout their entire range, there are slight diferences as shown on Figs. \ref{Fig2} ((d)) and \ref{Fig3} ((d)).

\section{Planar Wave Scattering}

In this section, we apply the partial-wave method to determine the scattering cross section associated with massless planar scalar waves interacting with the black hole. This approach allows us to obtain accurate numerical results that can be directly compared with the classical and semi-classical analyses discussed previously.

The Klein-Gordon wave equation for a massless scalar field propagating on a general gravitational background is given by;

\begin{equation}\label{Gordon}
\square \Phi=g^{\mu \nu}\nabla_{\mu} \nabla_{\nu} \Phi= \frac{1}{\sqrt{g}}\partial_{\mu}(\sqrt{-g}g^{\mu \nu}\partial_{\nu}\Phi)=0
\end{equation}

For a static spherically symmetric four dimensional space-time we shall consider monochromatic plane waves, on where;

\begin{equation}\label{field}
\Phi=\sum_{lm}\frac{\hat{\phi_l}(r)}{r}Y^{m}_{l}(\theta, \phi)e^{-i\omega t},
\end{equation}	

with $Y^{m}_{l}(\theta, \phi)$ the scalar spherical harmonics. By  introducing the radial partial wave functions $\Phi$ in (\ref{Gordon}) and after separation of variables we obtained the Regge-Wheeler equation

\begin{equation}\label{radial}
\frac{d^{2}\hat{\phi_l}(r)}{dr^{2}_{*}}+\omega^{2}\hat{\phi_l}(r) - V_{l}(r_{*})\hat{\phi_l}(r) =0.
\end{equation}

The variable $r_{*}= r_{*}(r)$ are the well known tortoise coordinates defined by the relation $dr_{*}/dr = 1/f(r)$ and $V_{l}(r_{*})$ is the Regge-Wheeler potential associated to the massless scalar field given by;

\begin{equation}
V_{l}(r)=f(r)\left(\frac{l(l+1)}{r^{2}}+\frac{f^{'}(r)}{r}\right)
\end{equation}

The scalar  potential is localized, going to zero at both limits $r_{*}\rightarrow \pm \infty$ ($r_{h}$ and $ \infty$). Then the waves coming from infinity are purely incoming from the past null infinity, obeying the following boundary conditions.

\begin{equation}\label{Front}
\hat{\phi_l}(r) = \left\{ \begin{array}{lr} A_1e^{-i\omega r_{*}} & r\to -\infty\quad\left(r\to r_h\right)\\ 
B_1e^{-i\omega r_{*}} + B_2 e^{i\omega r_*}  &  r\to \infty\quad\left(r\to \infty\right) \end{array} \right. 
\end{equation}

According to the assumed time dependence, the coefficient $A_{1}$ corresponds to the amplitude of the mode ingoing at the black hole horizon and the coefficient $B_{1}$ and $B_{2}$ corresponds to the amplitude of the mode incoming and outgoing respective from infinity. 

Then, it is possible to define the reflection and transmission coefficients as follows:

\begin{equation}\label{key}
R_{\omega}=\left|\frac{B_2}{B_1}\right|^2,\quad T_{\omega}=\left|\frac{A_1}{B_1}\right|^2,\quad\text{with}\quad
R_{\omega}+T_{\omega}=1
\end{equation} 

Then the scattering cross section for a static, spherically symmetric space-time, expressed in terms of partial waves, is given by:

\begin{equation}\label{key}
\frac{d\sigma}{d\Omega}=\left|g\left(\theta\right)\right|^2
\end{equation}

where $g(\theta)$ is the scattering amplitude given by;

\begin{equation}\label{g}
	g\left(\theta\right)= \frac{1}{2i\omega}\sum_{l=0}^{\infty}\left(2l+1\right)\left[S_{l}(\omega)-1\right]P_l\left(\cos\theta\right)
\end{equation}

where $S_{l}(\omega)$ is called the S-matrix defined as;

\begin{equation}\label{key}
S_{l}(\omega)= e^{2i\delta_l\left(\omega\right)}=
\left(-1\right)^{l+1}\frac{|B_2|}{|B_1|}.
\end{equation}

From  (\ref{g})  it is possible to observe that $g\left(\theta\right)$ exhibits singular behavior at $\theta=0$, which leads to convergence issues in the partial-wave series. In particular, as $ \theta \to 0 $, the singularity becomes dominant. To mitigate this problem, the function $ g(\theta) $ can be modified to soften its behavior at small scattering angles, as has been applied in the context of fermion scattering in the Schwarzschild space-time (see \cite{Dolan:2006vj}). This is achieved by multiplying $ g(\theta) $ by a factor that vanishes at $ \theta = 0 $. Accordingly, we can rewrite (\ref{g}) as:

\begin{equation}
\left(1-\cos\theta \right)^m 2i\omega\,g(\theta)
=\sum_{l} a_l^{(m)}\,P_l\!\left(\cos\theta\right)\,.
\end{equation}
The coefficients $ a_l^{(m)} $ are obtained recursively through the Legendre polynomial relations,
\begin{equation}
a_l^{(i+1)} = a_l^{(i)}
- \frac{l+1}{2l+3}\, a_{l+1}^{(i)}
- \frac{l}{2l-1}\, a_{l-1}^{(i)} \,.
\end{equation}

In this work, we adopt $ m = 2 $, which provides an optimal balance between convergence improvement and numerical stability for the present calculations. The results are also obtained by imposing the boundary conditions given in (\ref{Front}). The near-horizon condition is implemented at $r = r_{h} + 10^{-5}$, while numerical infinity is set at $r = 500$ considering $l=40$.

\begin{figure}[ht]
\begin{center}
\includegraphics [width =0.45 \textwidth ]{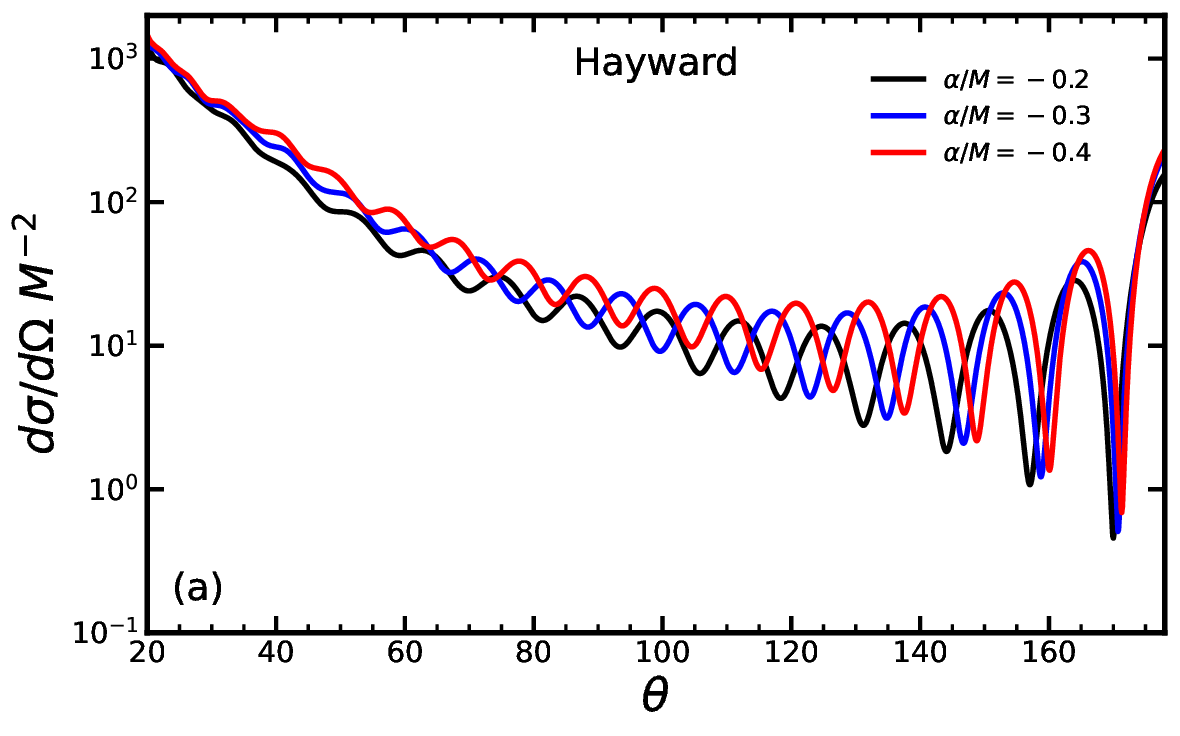}
\includegraphics [width =0.45 \textwidth ]{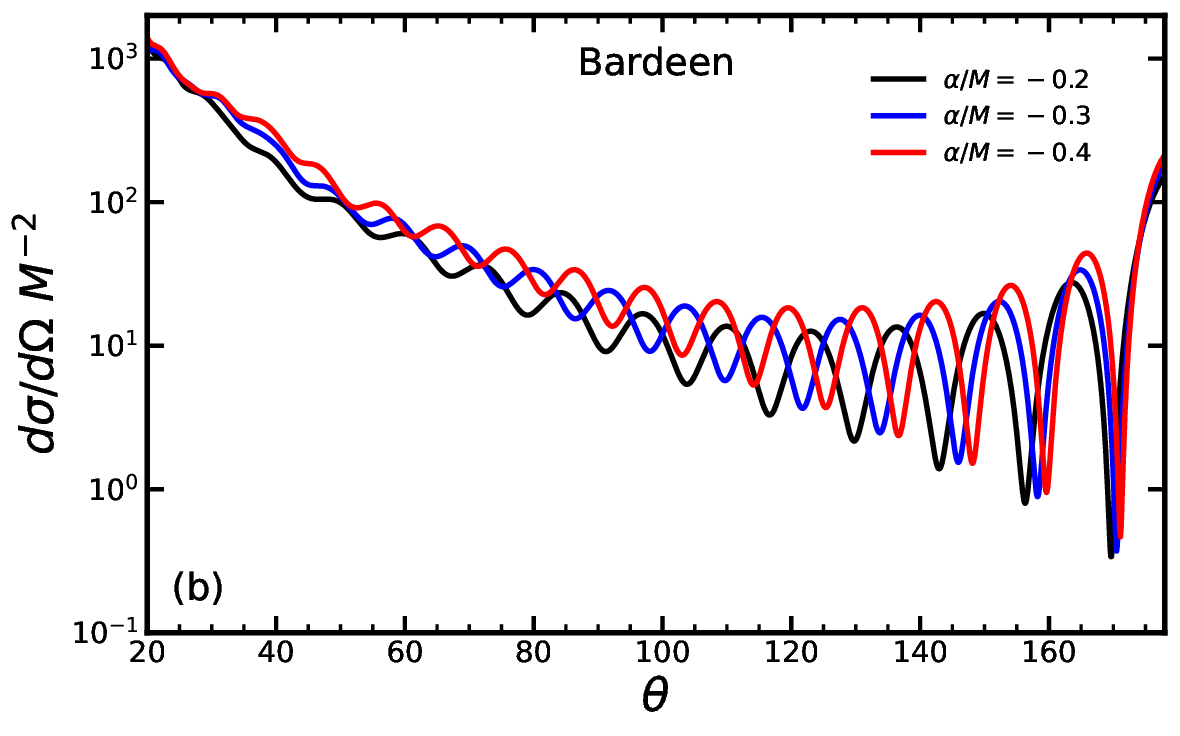}
\includegraphics [width =0.45 \textwidth ]{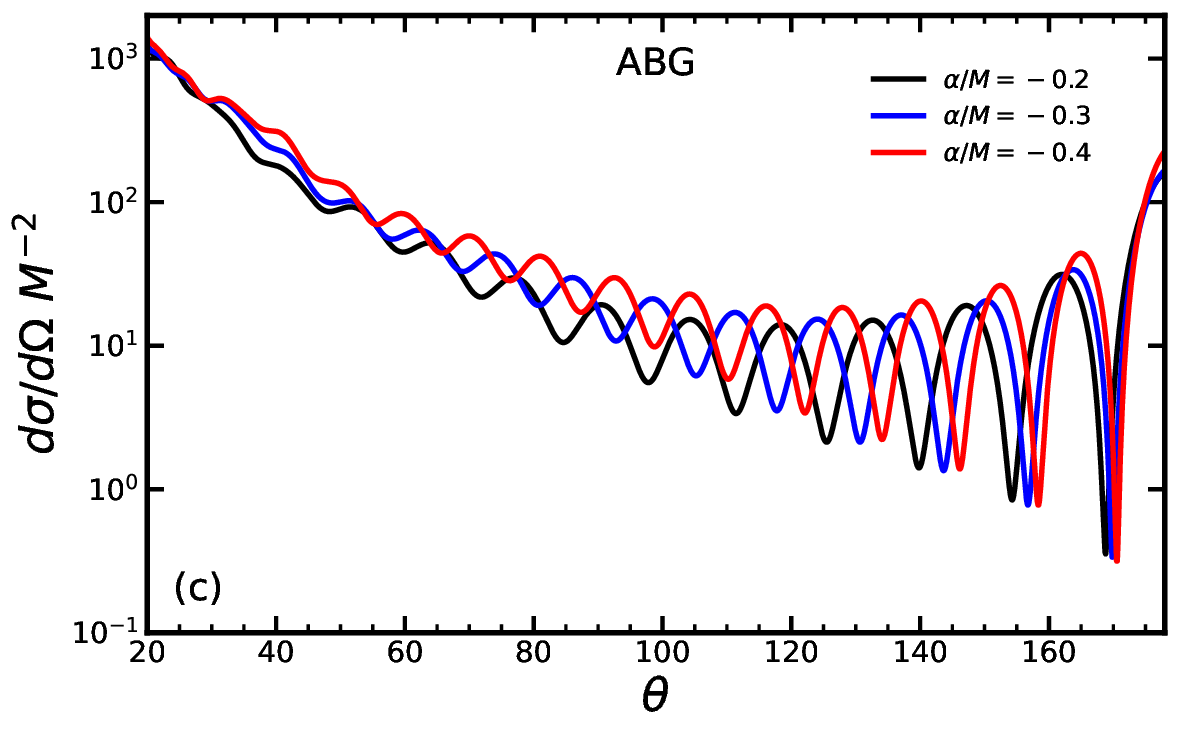}
\includegraphics [width =0.45 \textwidth ]{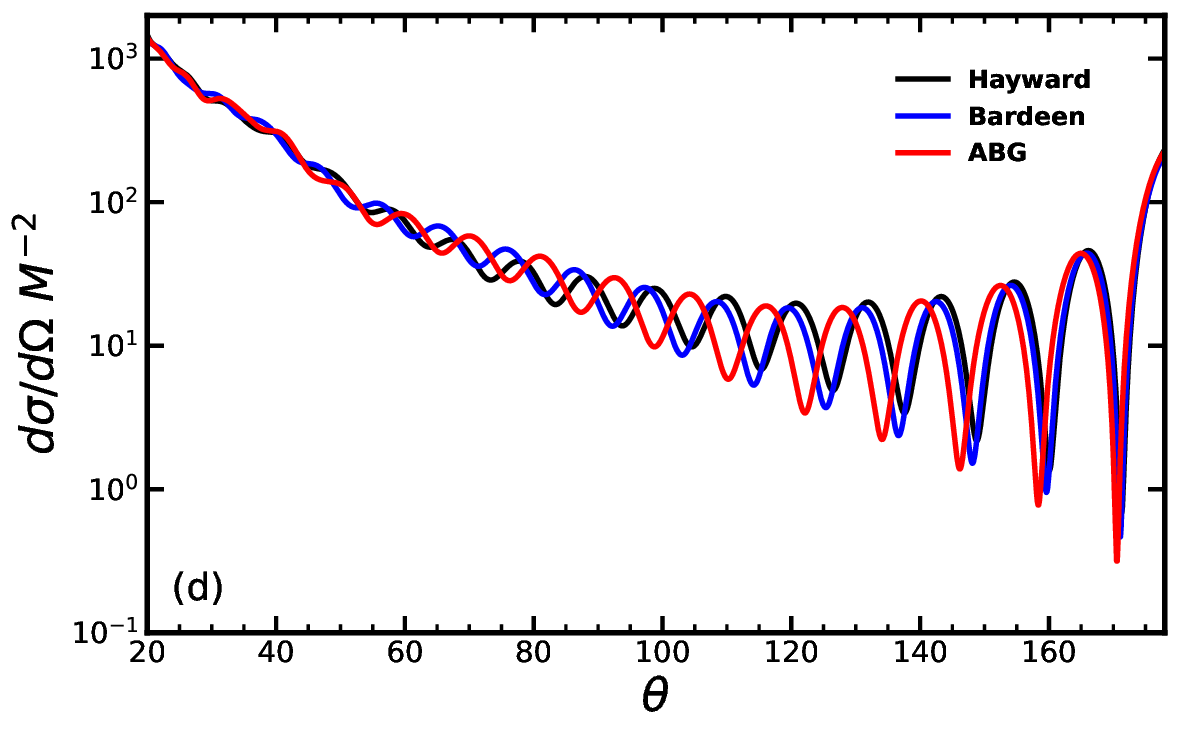}
\end{center}
\caption{Scattering cross sections for black holes immersed in PFDM: (a) Hayward black hole, (b) Bardeen black hole and (c) Ay\'on–Beato–Garc\'ia black hole are shown for different values of $\alpha /M$. Panel (d) presents a comparative analysis of the scattering cross sections for the BHs with $\alpha / M = -0.4$. For the different panels, we consider $(\epsilon/M)^{2}=(g/M)^{2}=(q/M)^{2}=0.6$ and $M\omega=2$}\label{Fig4}
\end{figure}

In Fig. \ref{Fig4}, we present the scattering cross sections for the Hayward black hole (Fig. \ref{Fig4} (a)), the Bardeen black hole (Fig. \ref{Fig4} (b)), and the ABG black hole (Fig. \ref{Fig4} (c)) immersed in PFDM, for different values of the parameter $\alpha$, obtained using the partial-wave method. As previously discussed, these results are derived by matching the numerical solution of the radial equation (\ref{radial}) to its corresponding asymptotic behavior. 
We observe that the widths of the interference fringes decrease as the contribution of dark matter increases, in agreement with the expectations from the semi-classical analysis.

Fig.~\ref{Fig4}(d) presents a comparison of the numerical scattering cross sections for the different BH models at a fixed value of the parameter~$\alpha$. Overall, the three configurations exhibit a very similar scattering behavior across the full angular range, with only minor deviations.

\begin{figure}[ht]
\begin{center}
\includegraphics [width =0.45 \textwidth ]{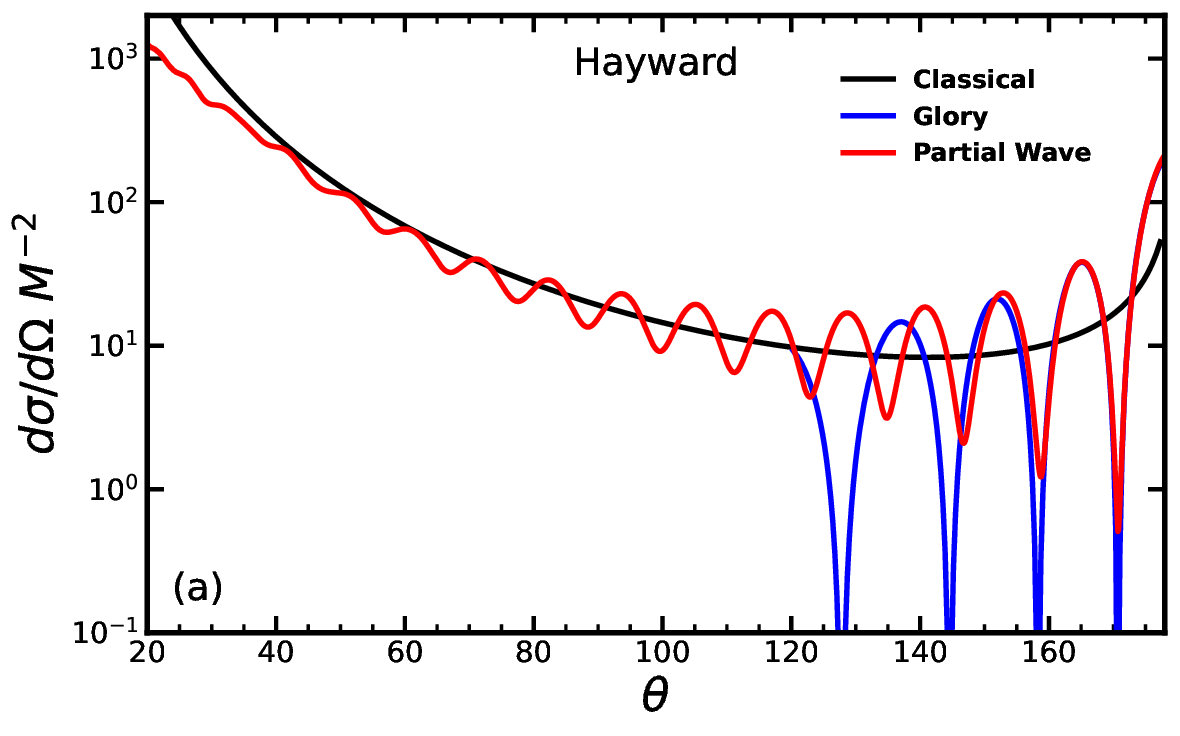}
\includegraphics [width =0.45 \textwidth ]{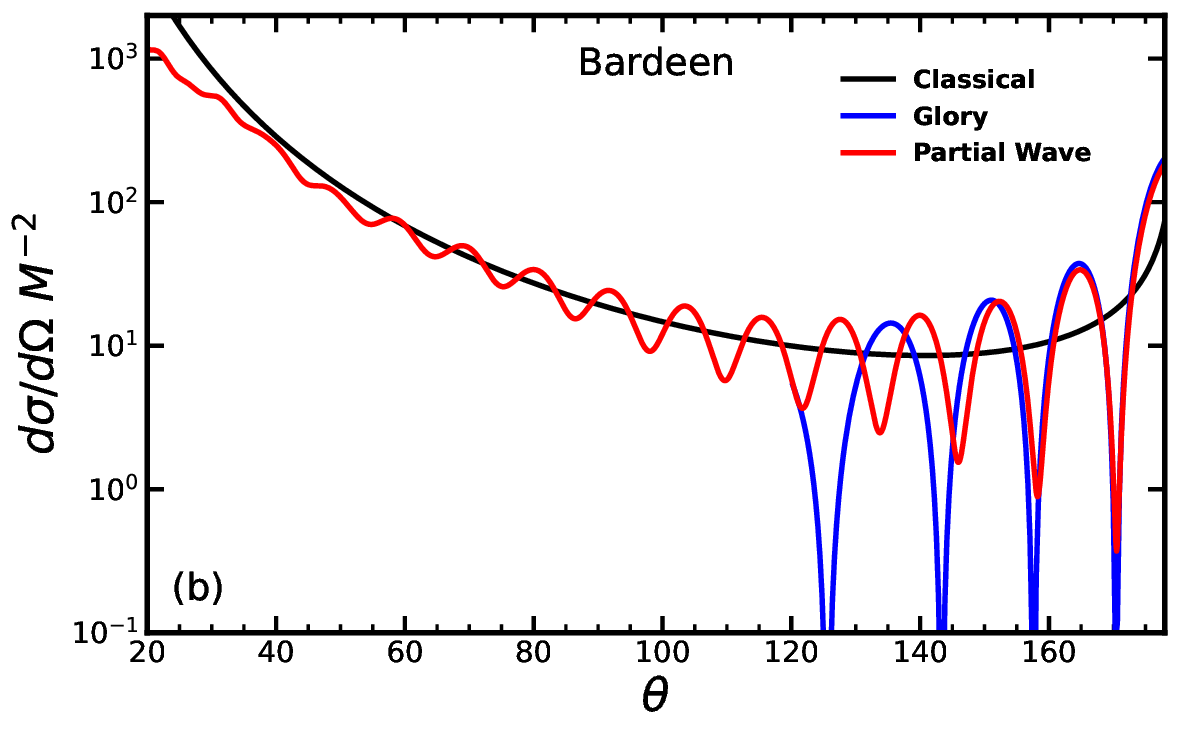}
\includegraphics [width =0.45 \textwidth ]{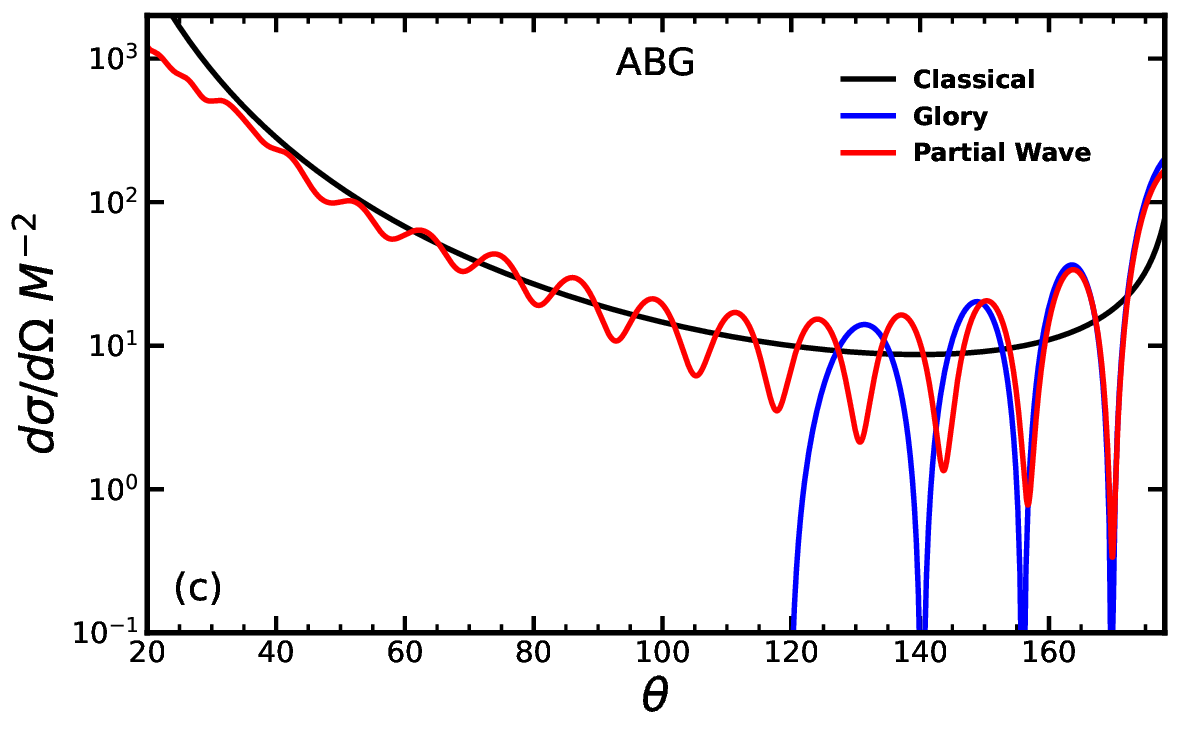}
\end{center}
\caption{Scattering cross sections for black holes immersed in PFDM: (a) Hayward black hole, (b) Bardeen black hole and (c) Ay\'on–Beato–Garc\'ia black hole, shown for $\alpha/M=-0.3$, $M\omega=2$ and $(\epsilon/M)^{2}=(g/M)^{2}=(q/M)^{2}=0.6$.}\label{Fig5}
\end{figure}

Finally, Fig.\ref{Fig5} shows the comparison of the scattering cross sections obtained via the partial-wave method, together with the classical geodesic approach and the glory approximation, for the different black holes with $M\omega=2$, while keeping the remaining parameters fixed. 

In all cases, we observe that the glory approximation corresponds well with the partial-wave results for scattering angles $\theta \gtrsim 140^{o}$, while the classical prediction accurately reproduces the behavior in the small-angle regime. This agreement also serves as a consistency check of our calculations, supporting the reliability of the results used to assess how the presence of dark matter modifies the scattering cross sections.

\section{Conclusions}

In this work, we analyzed the scattering cross sections of the Hayward, Bardeen, and Ay\'on--Beato--Garc\'ia regular black holes immersed in perfect fluid dark matter. First, the parameter space study shows that in presence of PFDM, the Hayward solution exhibits the widest region in which two horizons are allowed.

For the scattering analysis, we computed the classical, semi-classical, and partial-wave scattering cross sections for different values of the parameter $\alpha$. In all cases, the presence of PFDM enhances the magnitude of the scattering cross section. The semi-classical results further indicate that the interference fringes become narrower as $|\alpha|$ increases, reflecting the modifications introduced by the dark-matter term in the effective potential.

The partial-wave calculations confirm the trends obtained in the classical and semi-classical regimes and provide a consistent quantitative description of the scattering process. For large scattering angles, the glory approximation shows good agreement with the partial wave results, while the classical description accurately reproduces the small angle behavior. This consistency supports the reliability of the numerical procedures employed.

The results presented here extend previous analyses \cite{Tovar:2025apz} of regular black holes immersed in PFDM and contribute to the understanding of how dark matter environments modify scattering phenomena. 

\section*{ACKNOWLEDGMENT}

The authors acknowledge to SECIHTI-SNII, Mexico.

\bibliographystyle{unsrt}

\bibliography{bibliografia}

\end{document}